\documentclass[aps,prl,twocolumn,nofootnotebib,preprintnumbers]{revtex4-1}
\usepackage{epsfig}
\usepackage{bm}
\usepackage{graphicx}
\usepackage{amssymb}
\usepackage{amsmath}
\usepackage{color}
\usepackage{ulem}
\usepackage[caption=false]{subfig}
\usepackage{slashed}
\newcommand{\nn}{\nonumber\\&&}

\newcommand{\kslash}{\not\hspace{-0.7mm}k}
\newcommand{\ben}{\begin{displaymath}}
\newcommand{\een}{\end{displaymath}}
\newcommand{\be}{\begin{equation}}
\newcommand{\ee}{\end{equation}}
\newcommand{\bea}{\begin{eqnarray}}
\newcommand{\eea}{\end{eqnarray}}

\newcommand{\eq}[1]{Eq.~(\ref{#1})}

\def\g{\gamma}
\def\e{\epsilon}
\def\o{\omega}

 \def\D{\Delta}

\def\m{\mu}

\def\e{\epsilon}
\def\L{\Lambda}
\def\G{\Gamma}

\def\t{\tau}
\begin{document}                                           
                                                                         
\title{\bf  
{Chiral Light Front Perturbation Theory and the  Flavor Dependence of the Light-Quark Nucleon Sea    }}

\author{  Mary Alberg$^{1,2}$,  Gerald A. Miller$^2$}

\affiliation{$^1$Department of Physics, Seattle University, Seattle, WA 98122, USA }

\affiliation{$^2$Department of Physics,
University of Washington, Seattle, WA 98195-1560, USA}

\date{\today}                                                        
                                                                                                                                                                                 
\begin{abstract}
 The  light-quark flavor dependence  of the proton sea has been of great interest for many years because of its close connection with 
  non-perturbative effects.  One hypothesis is that  this dependence  arises from the   pion cloud of the proton.    We  apply light cone perturbation theory and experimental constraints to   a chiral Lagrangian to compute   the relevant Fock-space components of the nucleon wave function  with well-defined uncertainties.  
 Existing experimental information regarding   the light flavor sea is studied,  and predictions  for  future experimental results are provided. Future  experiments have the ability to rule out  this hypothesis 
 and have profound implications for understanding the nucleon-nucleon force.

\end{abstract}

\maketitle     
\noindent
\newpage

                            
\section{I. Introduction}

Textbooks tell us that nucleons are composed of $u$ and $d$ valence constituent quarks, but this cannot be the whole story because the gluons inherent in QCD must generate quark-antiquark pairs. Thus one is led to the question: Do the pairs arise only from perturbative evolution at high momentum scales, or do they have a non-perturbative origin? A definitive answer would provide great help in understanding the nature of confinement and also fundamental aspects of the nucleon-nucleon force.  Perturbative QCD predicts a sea that is almost symmetric in light flavor. However, the discovery of the violation of the Gottfried sum rule told us that  $\bar{d}$ quarks are favored over  $\bar{u}$ quarks
\cite{Amaudruz:1991at} . This   highlighted the importance of the pion cloud of the nucleon~\cite{Thomas:1983fh,Henley:1990kw}. 

 To truly understand the sea   one needs to know more.  
Efforts to measure the ratio $\bar{d}/\bar{u}$ by the E866 collaboration \cite{Hawker:1998ty,Towell:2001nh} have been ongoing and continue with the SeaQuest experiment \cite{Aidala:2017ofy}.  
Theory 
is reviewed in \cite{Speth:1996pz,Chang:2014jba}.

The pionic contribution to the  nucleon sea is of special interest. Understanding the pion and its interaction with and amongst nucleons is a necessary step in 
 learning how QCD describes    the   existence of atomic nuclei.
 As a  nearly massless excitation of the QCD vacuum, 
  the 
pion is a harbinger of spontaneous symmetry breaking. 
The pion is associated with large distance structure of the nucleon ~\cite{Theberge:1980ye,Thomas:1981vc,Theberge:1981mq} and its exchange between nucleons    provides  the longest ranged component of the    strong force. The general field  of chiral perturbation theory~\cite{Bernard:1992qa,Bernard:1995dp}  is well-established. 
 
A non-perturbative sea arises from the pion cloud.  
We  aim to make definitive predictions of the pion cloud model by   providing a light-cone perturbation theory approach  capable of making   predictions with known uncertainties.
Previous  calculations have  noted  ambiguities related to the dependence of the pion-baryon vertex function on momentum transfer and on the possible dependence upon the square of the four-momentum  of intermediate baryons, and much discussion has ensued~\cite{Holtmann:1996be,Koepf:1995yh,Speth:1996pz,Strikman:2009bd,Strikman:2010pu,Alberg:2012wr,Ji:2009jc,Burkardt:2012hk,Ji:2013bca,Salamu:2014pka,Granados:2015rra,Granados:2016jjl}.

There is a vast, deep and venerated literature on the role of the pion cloud in nucleon structure. Nevertheless, there is still much to be done. For example, even the very first example~\cite{Thomas:1983fh,Henley:1990kw}, the size of the pion contribution to the $\bar d-\bar u$ asymmetry, is under present controversy. Refs.~\cite{Alberg:2012wr,Salamu:2014pka}, argue that the pion cloud accounts completely for the measured asymmetry. In contrast,  Ref.~\cite{Strikman:2009bd} finds that only about half of the asymmetry can be accounted for by pion cloud effects. In their view, cancellations between the effects of $N$ and $\D$ intermediate states, which would be complete in the limit of large $N_c$, reduce the size of the effect. This discrepancy needs to be resolved.

There is another more fundamental issue involving the loss of relativistic invariance which occurs when the vertex function is treated (universally in all of the models so far) as depending on only three of the four necessary momentum variables. This paper resolves both of these problems by using a four-dimensional formalism and by using experimental constraints on the pion-baryon vertex function. The formalism employed here combines pion-baryon dynamics with light front perturbation theory. Hence the new nomenclature, `chiral light-front perturbation theory' that appears in the title. 
   
 \section{II. Formalism}
  
 In a light cone perturbation theory (LCPT) description, the proton wave function can be expressed as a sum of Fock-state components
 \cite{Lepage:1980fj,Brodsky:1997de,Brodsky:2000ii,Kovchegov:2012mbw}.     Our   hypothesis is that  the non-perturbative light-flavor sea  originates from  
  the bare nucleon,   pion-nucleon ($\pi N$) and  pion-Delta ($\pi\D$) components.
   The interactions are described by  using the relativistic  leading-order chiral Lagrangian \cite{Becher:1999he,odgt, Pascalutsa:1998pw}. 
  Displaying 
  the interaction terms to the relevant order in powers of the pion field, we use
 \bea
\mathcal{L}_{\rm int}& = 
 - \; {g_A\over 2 f_\pi} \bar\psi \gamma_\mu \gamma_5 \tau^a \psi 
\, \partial_\mu \pi^a 
 - {1\over f_\pi^2} \bar\psi \gamma_\mu \tau^a \psi \;
\e^{abc} \pi^b \partial_\mu \pi^c \nonumber\\&-{g_{\pi N\Delta}\over 2M} (\bar{\Delta}^i_\mu g^{\mu\nu}\psi\partial_\nu \pi^i +{\rm H. C.}) 
\label{lag}
\eea
 where $\psi$ is the Dirac field of the nucleon,   $\pi^a (a = 1, 2, 3)$
the chiral pion field and $M$ is the nucleon mass. In Eq.~(\ref{lag})
$g_A$ denotes the nucleon axial vector coupling and $f_\pi$ the 
pion decay constant.  The second  term  is the Weinberg-Tomazowa term which describes  
low-energy $\pi-$nucleon scattering. In the third term $g_{\pi N\Delta}$ is the $\pi N\Delta$ coupling constant, and the $\Delta^i_\mu$ field   is a vector   in both spin and  isospin space. 

Previous work~\cite{Alberg:1999bc} included the effects of the $\o$ meson, 
based 
on one-boson exchange models of the NN 
potential. Current treatments 
do not include the explicit effects of  exchanged vector mesons as their masses represent a high-energy scale~\cite{Bedaque:2002mn,Machleidt:2011zz,Machleidt:2017vls}. 
When computing spin-independent quantities the small expected effects of such heavy mesons cannot be distinguished from 
  the bare sea of the nucleon.  The influence of heavy mesons may be important for spin-dependent effects (see the review~\cite{,Chang:2014jba})  and that topic will be discussed in a later paper.

 \begin{figure} [h]
\includegraphics[width=7.1991cm]{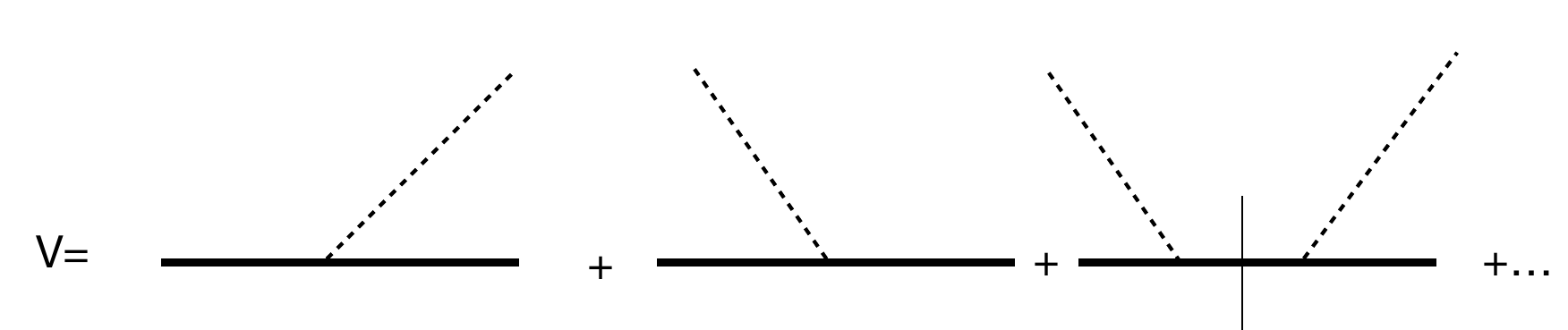}
\caption{Terms in the light-front Hamiltonian. }\label{Vpic}\end{figure}

    The 
    procedure for deriving the  LCPT for any Lagrangian is to construct the Hamiltonian operator  
   from the $T^{+-}$ component of the energy momentum tensor~\cite{Lepage:1980fj,Brodsky:1997de,Yan:1973qf,Miller:1997cr,Kovchegov:2012mbw}.The Hamiltonian can be written in terms of a sum of kinetic energy operators, $M_0^2$ and interaction  terms, denoted as $V$, see Fig.~\ref{Vpic}.  The first two terms are standard interactions, and the third is an instantaneous term  that     enters   only at 
   higher orders in the coupling constant.  The Hamiltonian forms of the  single-pion emission or absorption terms  (Fig.~\ref{Vpic}) are expressed as matrix elements evaluated between on-shell free nucleon spinors~\cite{Lepage:1980fj,Brodsky:1997de,Yan:1973qf,Miller:1997cr,Kovchegov:2012mbw}.  The light-front Schroedinger equation for the proton, $p$, is given by:
$ (M_0^2 +V)|p\rangle=M_p^2|p\rangle.$ 
To the desired second-order :
\bea |p\rangle\approx \sqrt Z |p\rangle_0+{1\over M_p^2-M_0^2}V|p\rangle_0,\label{wf}\eea
where $|p\rangle_0$ represents the  nucleon in the absence of the pion cloud, the bare nucleon, and $Z$ is a normalization constant. 
 This expression is of first-order in $V$, hence the  term `perturbative'. The model is based on the assumption that including  higher-order terms is not necessary because including higher-order diagrams would introduce large uncertainties. However,  as explained below, the consistency of this approach is maintained by using soft-form vertex functions.
Given \eq{wf}, the wave function can be  expressed as a sum of Fock space components given  by 
\bea
|p\rangle=\sqrt Z |p\rangle_0+ \sum_{B=N,\D}\int d\Omega_{\pi B}|\pi B\rangle \langle \pi B|p\rangle_0, \label{fsw}
\eea
where $\int d\Omega_{\pi B}$ is a phase-space integral~\cite{Brodsky:2000ii,Kovchegov:2012mbw}.
In this formalism the  pion momentum distributions $f_{\pi B}(y)$, which represent the probability that a nucleon will fluctuate into a pion of light front momentum fraction $y$ and a baryon of light front momentum fraction $1-y$, are   squares of wave functions, $\left|\langle \pi B|\Psi\rangle\right|^2$ integrated over $k_\perp.$

The Lagrangian of  \eq{lag}  is incomplete because it is not renormalizable. We  tame  divergences using a physically motivated set of regulators,  depending on 4-momenta, that are constrained by data.  
 If chiral symmetry  is maintained, 
 one finds that the $\pi N$ vertex function $ g_{\pi N}(t)$  and the nucleon axial form factor are related  by the 
generalized Goldberger-Treiman relation~\cite{Thomas:2001kw} (obtained with $m_\pi = 0$):
 \bea &
 Mg_A(t)=f_\pi g_{\pi N}(t),\label{relate}
\\&g_A(t)=g_A(0)/(1+(t/M_A^2))^2\label{MA} \eea where $t$ is the square of the four-momentum transferred to the 
 nucleon.  \eq{relate} follows from  partial conservation of the axial-vector current  (PCAC)  and the pion pole dominance of the pseudoscalar current and  is obtained from  a matrix element of the axial vector current between two on-mass-shell nucleons.  
 The $t$-dependence of $g_A$ is determined for $t>0$ by low-momentum transfer experiments~\cite{Thomas:2001kw}, with $M_A$  the single parameter.   
   \eq{relate}   relates an essentially unmeasurable quantity $g_{\pi N}(t)$ with one $g_A(t)$ that is constrained by experiments.  The major uncertainty in previous calculations is largely removed. Some models, see e.g. \cite{Guichon:1982zk}, find differences between the $t-$dependence of $g_A(t)$ and $g_{\pi N}(t)$, which is allowed because $m_\pi \ne 0$.
   Uncertainties in the parameter $M_A$ are discussed below, where it is also shown that very large values of $t$  are not important in the calculations  of this paper.
   
In evaluating the nucleon wave function \eq{fsw} the necessary vertex function must be applicable to situations when either pion or baryon or both are off their mass shells.
 We use      frame-independent   pion-baryon form factors, in which a nucleon of mass $M$ and momentum  $p$ emits a pion of mass $\m$ and momentum $k$ and becomes a baryon of mass $M_B$ and momentum $p-k$:
 \bea &F(k , p, y)={\L^2 \over  k^2-\mu^2 -\L^2 +  i \e}\, {\L^2 \over {y\over 1-y}((p-k)^2-M_B^2)-\L^2+  i \e}. \label{BigF}\eea
 
   
 Using   $F(k,p,y)$ allows us to obtain a  pion-baryon  light front wave function. We explain for  the $\pi N$ system. The form factor is  part of a model for the $\pi N$
 vertex function $\G_a(k,p)=\kslash\g^5{g_A\over 2 f_\pi}\tau_a F(k,p,y)$, where $k$ is the pion momentum and $p$ the nucleon momentum. The $\pi N$ component of the light-front wave function is obtained from the Bethe-Salpeter wave function, $\Psi_{\pi N}^a(k,p)$ by integrating over $k^-$~\cite{Lepage:1980fj,Carbonell:1998rj,Miller:2009fc} and projecting onto on-mass-shell baryon spinors:
 \bea 
 \Psi_{a{\rm LF}}(k,p,s)={M^2 y(1-y)\over 2\pi p^+}\int_{-\infty}^\infty dk^-\Psi_{\pi N}^a(k,p,s), \nonumber\eea
with $\Psi_{a{\rm LF}}(k,p,s) \equiv \langle k,p,s| \pi N\rangle $ and  
\begin{widetext}
\bea  
\Psi_{\pi N}^a(k,p,s) = -i \bar{u}(p-k,s)\kslash\g^5u(p,s){g_A\over 2 f_\pi}\tau_a  {1\over (k^2-\mu^2+i\e)((p-k)^2-M^2+i\e))}F(k,p,y).
\eea
\end{widetext}
 The definition of the  light-front wave function uses the basis that the Dirac spinors are evaluated for  on-shell kinematics  with  
$(p-k)^-= ((p-k)_\perp^2+M^2)/ (p-k)^+$, 
 so that the numerator factor does not depend on $k^-$, and the  integration over $k^-$ involves only the denominator. 
The evaluation of   $\Psi_{a{\rm LF}}(k,p,s)$ by integrating over the upper half $k^-$ plane   
gives the same result as  integration over the lower half. Thus the light-cone wave function (including the effect of form factors) is uniquely defined.  Both procedures yield:
\bea
& \Psi_{\rm a,LF}(k,p,s) 
 ={M g_A\over 2f_\pi (2\pi)^{3/2}}\sqrt{{y}\over  1-y} { \bar{u}(p-k)i\g^5\tau_a u(p)\over t+\mu^2} F_A(t),\nonumber\\&F_A(t)\equiv{2 \L^4\over \left(\L^2+t+\mu ^2\right) \left(2\Lambda ^2+t+\mu ^2\right)}.\label{newr}
\eea
 Expanding $F_A(t)$ to first order in $t$, comparing the result to the same expansion of  $g_A(t)/g_A(0)$ and matching the results determine the value  $\L=\sqrt{3}/2M_A$. Numerical results (see below) show  that  using  this $F(k,p,y)$  is equivalent to using a form factor of the form of \eq{relate}  in computing $f_{\pi N}(y)$.  The parameter-independence of this approach is maintained.\\

The pion 2D momentum distribution function $f_{\pi N}(y,t)$ is obtained by squaring  $\left|\Psi_{\rm a,LF}(k,p,s)\right|$ and summing over $a,s$. The result is
\bea  { f_{\pi N}(y,t)={3M^2\over 16\pi^2} {g_A^2\over f_\pi^2} y  {t\over(t+\m^2)^2} F_A^2(t)},
\label{nlf1} 
\eea
with  $t = (M^2y^2+k_\perp^2)/(1-y)$. 
The pion longitudinal momentum distribution function $f_{\pi N}(y)$ is then                    
\bea
f_{\pi N}(y) = \int_{t_N}^{\infty}{dt }f_{\pi N}(y,t)
\eea
 where $t_N\equiv{M^2y^2/(1-y)}$.
    Using  \eq{relate} or \eq{newr}  yields the same  $f_{\pi N}$ 
  because the integrand is dominated by the region of low values $t$, see below, Fig. 6.
  
   It is necessary to show that the pionic effects are of long range when  the stated vertex function is used. The formal way to do that is to study the resulting  three-dimensional, light-front  structure of the pion-baryon component. 
   Here
  the transverse spatial probability density of the $\pi N$ fluctuation,  $\rho_{\pi N}(y,b) = | \psi_{\pi N}(y,b)|^2,$, with 
  
  \bea   
  \psi_{\pi N}(y,b) = { 1\over (2 \pi)^2} \int_{0}^{\infty}{d^2 \vec{k}_\perp e^{i \vec{k}_\perp\cdot \vec{b}}  \Psi_{\rm a,LF}(k,p)}
 \label{rhoyb} 
\eea 
with $k_\perp = \sqrt{(1-y)t-y^2M^2}$.
  This distribution represents the spatial extent of the pion cloud, and as pointed out by Burkardt \cite{Burkardt:2002hr}, there is no direct connection between transverse momentum space densities $f_{\pi N}(y,k_\perp)$ and transverse position space densities  $\rho_{\pi N}(y,b)$.

Next is  the   intermediate $\D$ contribution. 
The pionic coupling between nucleons and $\Delta$ particles has an off-diagonal Goldberger-Treiman relation \cite{odgt,arXiv:0706.3011}  similar to  \eq{relate}.
Lattice calculations \cite{arXiv:0706.3011}
show that ${g_{\pi N\Delta}(t)\over g_{\pi N}(t)}=1.61$  
is constant  
and consistent with the Goldberger-Treiman relations. 

The evaluation proceeds 
as for the intermediate nucleon.  
 The result is 
 \bea&& 
  f_{\pi\Delta}(y)={1\over 12 \pi^2} \left({ g_{\pi N\Delta}\over 2M}\right)^2  y
\int_{t_\D}^\infty  dt\; {\;F_A^2(t)\over (t+\mu^2)^2}\nn\times
\left (t+{1\over4M_\Delta^2}(M^2-M_\Delta^2+t)^2\right)
((M+M_\Delta)^2+t) \nonumber\\&&\label{piD}
\eea
with $t_\D=(y^2M^2+y(M_\Delta^2-M^2))/(1-y)$.  
 

The previous material completes the discussion of the formalism. 

\section{III. Application to parton distributions}

\begin{figure}[h]
\includegraphics[width=4.291cm]{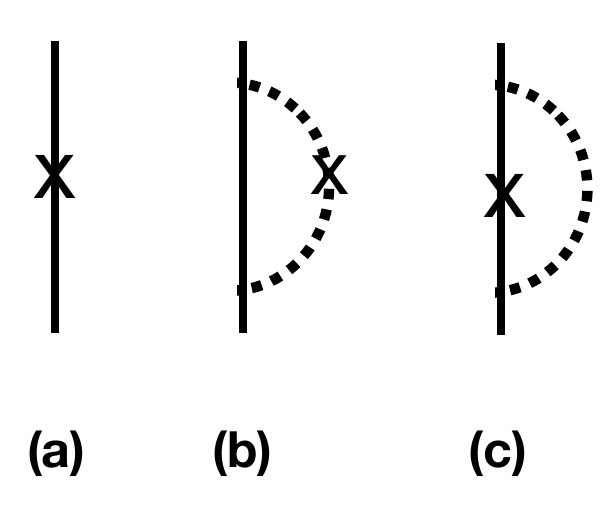}
\caption{(a) External interaction, X, with bare nucleon (solid line) (b)  External interaction, X, with the pion (c) External interaction, X, with the intermediate baryon. Here X represents the deep-inelastic scattering operator. }\label{fig:diag}\end{figure}

The next step is
to  
  use \eq{fsw} to compute the  light flavor sea 
  of a nucleon. Consider the role of the pion cloud in 
deep inelastic scattering,  (Fig.~\ref{fig:diag}).    One needs to include   terms in which the virtual photon hits (a) the bare nucleon, (b) the intermediate pion~\cite{Sullivan:1971kd}   and (c)  the intermediate baryon   $B$  of the $(\pi B)$ Fock-state component. 

The effects of the Weinberg-Tomazowa (WT)  term  vanish    because the  deep-inelastic scattering (DIS) operator, represented by  X in the figure is diagonal in the pion flavor index. Deep inelastic scattering is a form of virtual Compton scattering. This scattering operator only takes a pion of a given charge into a pion of the same charge. This operator cannot change a flavor index so there is no effect. In more detail, denoting the DIS operator as ${\cal M}_{bb}$ the evaluation of the WT looks schematically like $\t^a\e^{abc}\pi^b\partial_\m\pi^b {\cal M}_{bb}=0$. Of course, the electromagnetic interaction, which explicitly depends on the charge of the pion and is off-diagonal in the flavor indices, does yield a contribution as has been known for a long time, see {\it e.g.} \cite{Burkardt:2012hk}. However, any implication that the WT term leads to a contribution to DIS is simply wrong.
 Another key assumption of the present model is that   quantum interference effects involving different Fock space components are negligible
  because different final states obtained from deep  inelastic scattering  by the pion and by the nucleon are expected to be orthogonal.

Given the lack of interference effects, 
  one can represent the  quark distribution functions of flavor $f = (\bar{u},\bar{d})$ in the nucleon sea as
\bea &q^f_N(x)=Z q_{N0}^f(x)+\sum_{B=N,\D} f_{\pi  B}\otimes q_\pi^f +\sum_{B} f_{B\pi}\otimes q_B^f,\nonumber\\\label{convol}\eea
 in which 
$ f_{\pi B}\otimes q_\pi^f \equiv \int_x^1{dy\over y} f_{\pi B}(y)q_\pi^f({x\over y}).$ 
 The first symbol in the subscript represents the struck hadron, and the phase space factor in \eq{fsw} ensures that $f_{\pi B}(y)=f_{B\pi}(1-y)$.
The quark distributions of the hadrons in the cloud are given by $q_\pi^f(x)$ and $q_B^f(x)$, and the bare nucleon distributions are given by $q_{N0}^f(x)$. We assume that the  $\bar{u},\bar{d}$ parton distributions of the bare proton and  the intermediate $\D$ and nucleon are the same, $q^f_{N0}=q^f_\D=q^f_B$, and are flavor symmetric, as expected from the  perturbative generation of sea quarks
 via the flavor--independent quark-gluon coupling constant.

 Our Fock space  expansion  has no  one-to-one correspondence with Feynman diagrams. For example, the Feynman diagram of Fig.~2b would  involve transitions from two-pion states to zero pion states. Such effects 
 should not be included  because the light-cone Fock space wave function is used to obtain probability distributions. 
  
Contributions to the antiquark sea of the proton come from the valence  and sea distributions of the pion $q_\pi^v$   and $q_\pi^s$
and the sea distributions $q_B^s$ and $q_N^s$ of the intermediate baryons and the bare proton. 
 The use of these distributions to describe deep inelastic scattering from a bound pion  follows from the light-front Fock space expansion, \eq{fsw}, which involves  only  on-mass-shell constituents. 
  With 
$
f_{\pi^+ n}={2\over 3}f_{\pi N},\: f_{\pi^0 p}={1\over 3}f_{\pi N},\:
f_{\pi^- \D^{++}}={1\over 2}f_{\pi \D},\: f_{\pi^0 \D^{+}}={1\over 3}f_{\pi \D},\:
 f_{\pi^+ \D^{0}}={1\over 6}f_{\pi \D},
$
the antiquark distributions are
\bea
&\bar{d}(x)= ({5\over 6}f_{\pi N}+{1\over 3}f_{\pi \Delta})\otimes q_\pi^v+ \bar{q}_{sym}(x)  \label{bard}\\
&\bar{u}(x)= ({1\over 6}f_{\pi N}+{2\over 3}f_{\pi \Delta})\otimes q_\pi^v+ \bar{q}_{sym}(x) \label{baru}\eea
where $ \bar{q}_{sym}(x)\equiv  \sum_ {B} f_{\pi B}\otimes q_\pi^s+\sum_ {B} f_{B \pi}\otimes q_B^s+ Z q_N^s(x).$
 The ${\pi N}$ terms favor the $\bar{d}$, but the ${\pi \D}$ terms favor the $\bar{u}$.\\

The pion  valence  quark parton distribution function (pdf) is obtained by  evolving  the pion valence pdfs of Aicher, Sch\"{a}fer and Vogelsang (ASV) \cite{Aicher:2010cb} from their starting scale to $Q^2 = 54$ GeV$^2$, which is the scale relevant for  E866. The present  fit to the evolved valence distribution
is given by
$q_\pi^v(x) =1.39\, x^{-0.331} (1-x)^{3.12 } \left(7.18\,  x^2+1\right).$ 
As in the ASV analysis,  the  pion sea quark pdfs  are  those   of Gluck, Reya and Schienbein  \cite{Gluck:1999xe}, for which at $54$ GeV$^2$
$q_\pi^s(x) = 0.115\, x^{-1.21} (1-x)^{5.34} \left(1-2.38\,
   \sqrt{x}+4.28\, x\right).$

Holtmann et al.~\cite{Holtmann:1996be} explained that the bare proton sea cannot directly  be determined from experimental data, which includes contributions from the pion cloud. They ultimately  \cite{Szczurek:1997fr}  used a fit to DIS data that included corrections for the pion cloud to determine the bare proton sea. We use their symmetric sea:
$x q_{N0}^{\bar d}(x) =x q_{N0}^{\bar u}(x) = {0.217 (1-x)^{15.6} (1 + 0.625 x)}.$ 
This distribution is also used   for terms  of Fig. 2(c). 


 Other input parameters must be described before presenting numerical results. The pion-nucleon splitting function $f_{\pi N}(y)$ depends on the   coupling constant $g_{\pi N}$ and the form factor cutoff $\L$. The lower limit for $g_{\pi N}$  is 12.8, taken from the Goldberger-Treiman relation $g_{\pi N} ={M\over f_\pi}g_A$, with $g_A=1.267\pm0.04,\;M=0.939\; {\rm GeV},f_\pi=92.6\;{\rm MeV}$.   The upper limit is $g_{\pi N} =13.2$, consistent with the scattering data analysis of Perez et al. \cite{Perez:2016aol} and the muon-based determination of $g_A$ by Hill et al. \cite{Hill:2017wgb}.   As noted above the cutoff parameter of \eq{newr},  $\L=\sqrt{3}/2M_A$, is obtained at very low $t$.  The two resulting splitting functions are identical for all values of $y$, demonstrating that only small values of $t$ are important in the present calculations. In initial calculations we used the value
  $M_A = 1.03 \pm 0.04$ 
 GeV~\cite{Thomas:2001kw}. This early review result  was  confirmed by many authors~\cite{Bernard:2001rs,Juszczak:2009qa,Katori:2016yel,Nakamura:2016cnn,Meyer:2016oeg}, all obtaining results within the  stated  uncertainty. We have increased the uncertainty in our cutoff $\Lambda$ to $\pm 10\%$ to allow for a difference between the cutoffs in the $\pi NN$ form factor and the axial form factor. Although one early estimate, based on the cloudy bag model, suggested that the difference might be $\pm 20\%$ \cite{Guichon:1982zk},  later work using dispersion relations found consistency between the axial form factor cutoff and a  $\pi NN$ {\it monopole} cutoff
   of $\Lambda =0.80$ GeV $\pm 10\%$  \cite{Bockmann:1999nu,Ericson:2000md}. A monopole value of $\Lambda= $ 0.8 GeV corresponds to a dipole value of 1.1 GeV. 
 
 \section{IV. Numerical results}
 
\begin{figure}
\includegraphics[width=8.0cm]{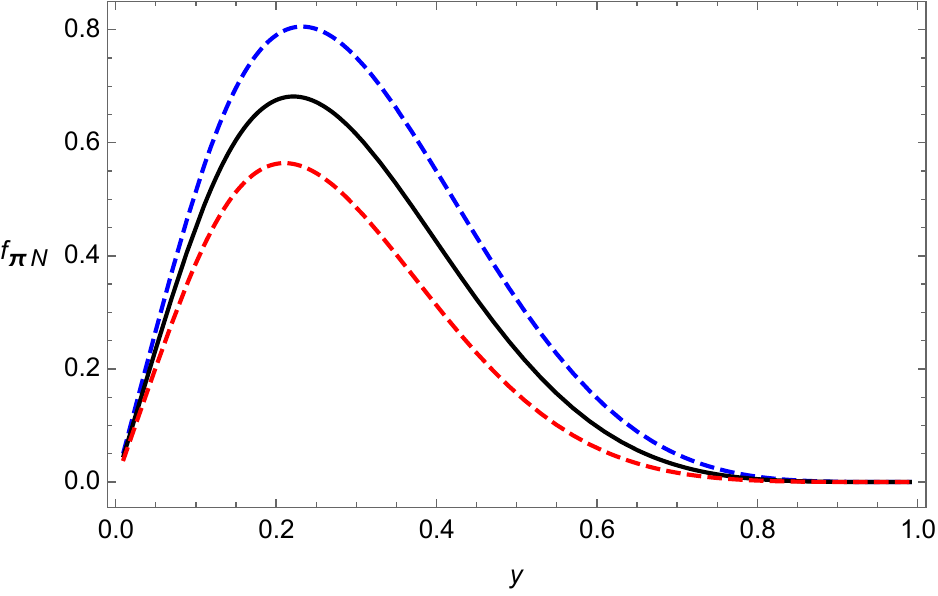}
\includegraphics[width=8.0cm]{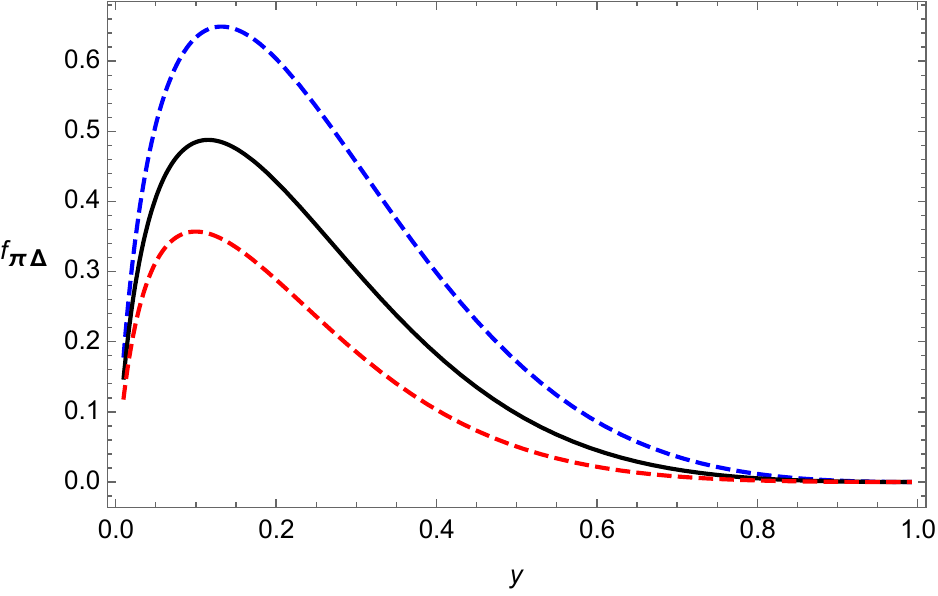}
\includegraphics[width=8.0cm]{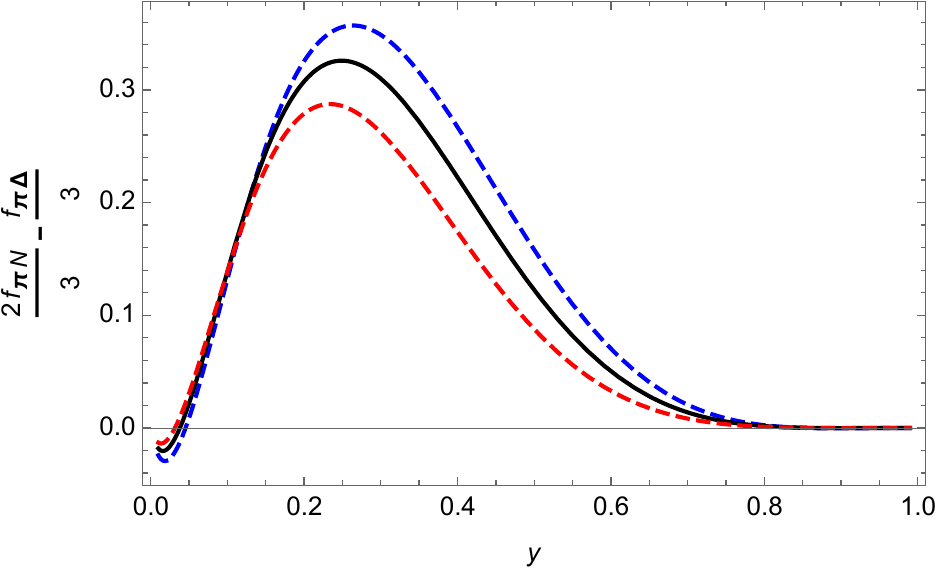}
\caption{Pion baryon splitting functions $f_{\pi {B}}(y)$, $B=N,\D$, are shown in the upper two panels. The solid lines are found using the central values of our coupling constants and cutoffs. The upper (blue) and lower (red) dashed lines are obtained using the maximum or minimum values, respectively,  of these parameters. The lowest panel shows the contribution of the splitting functions to the integrated asymmetry, $\bar{D}-\bar{U}$, \eq{intasym}. The smaller spread between the dashed lines is due to the correlation between the coupling constants and the use of the same cutoff in  $f_{\pi {N}}(y)$ and $f_{\pi {\D}}(y)$. }
\label{fig:splitfcns2}
\end{figure}

     The splitting function  $f_{\pi N}(y) $ is shown in  Fig.~\ref{fig:splitfcns2}  for a range of parameters bounded by the maximum and minimum values of $g_{\pi N}$ and $\L$.
The splitting function $f_{\pi \D}(y)$ depends on the coupling constant $g_{\pi \D}$ and the form factor cutoff $\L$.   We use the same form factor and cutoff for $f_{\pi N}(y)$ and $f_{\pi \D}(y)$. The upper limit of the coupling constant is obtained by using the quark model result
$
({g_{\pi \D} \over g_{\pi N}})^2= {72\over 25}, 
\; g_{\pi \D}=1.7 g_{\pi N}.
$ The lower limit of the coupling constant is obtained from the large $N_C$ limit of $g_{\pi \D}=1.5 \, g_{\pi N}.$
The ratio $f_{\pi \D}(y)/f_{\pi N}(y) $ { is less than unity for the important regions of $y$. It does  increase as $y$ increases above 0.5, and becomes greater than unity at about $y=0.8$, where both splitting functions are vanishingly small.}
 \\

%
%

\begin{figure}
\includegraphics[width=8.0cm]{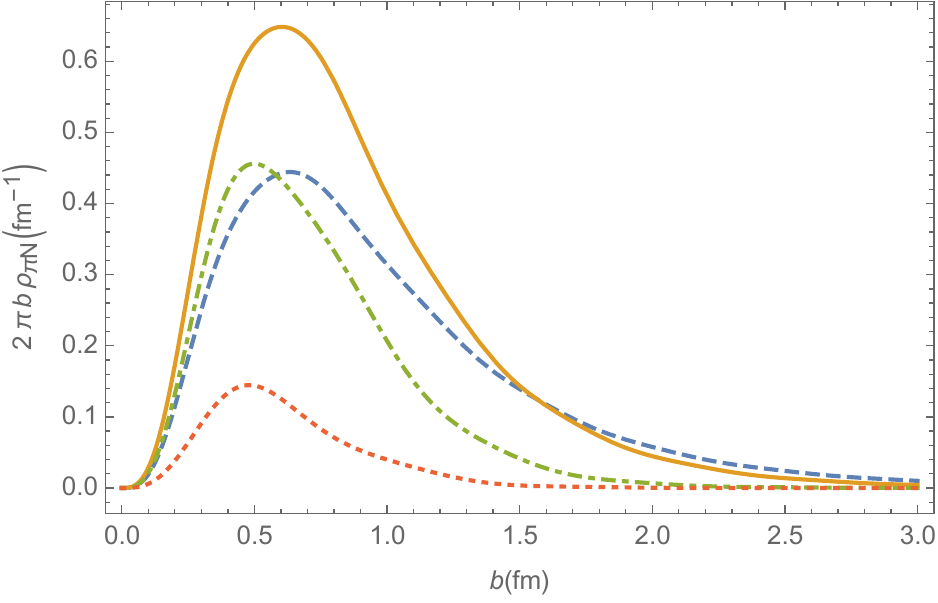}
\caption{$2\pi b\, \rho_{\pi N}(y,b)$ for $y$= 0.1 (dashed), 0.2 (solid), 0.4 (dot-dashed) and 0.6 (dotted), using the central values of our parameters 
for  $f_{\pi N}(y)$.}
\label{fig:brho}
\end{figure}

\begin{figure}
\includegraphics[width=8.0cm]{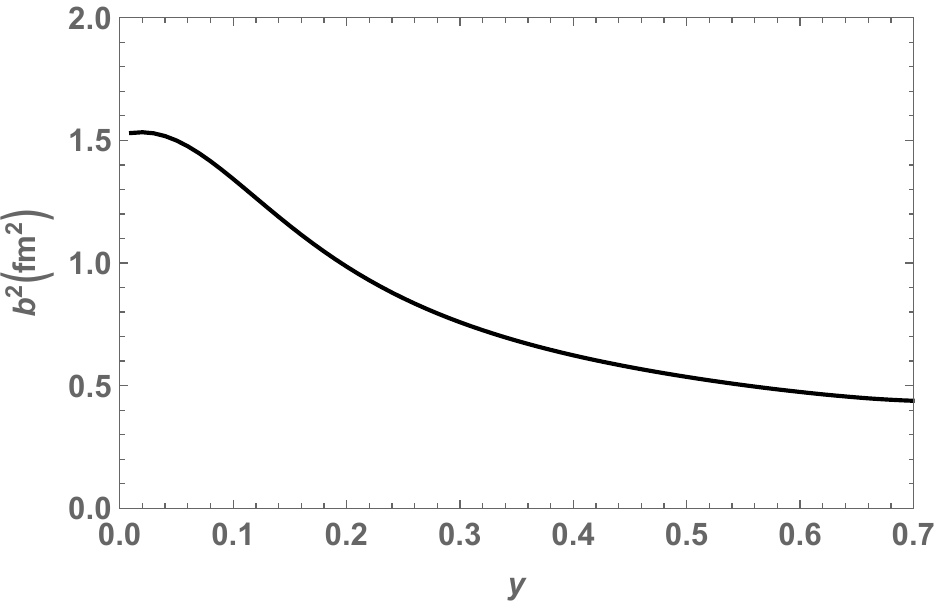}
\caption{$<b^2(y)>$ for $\rho_{\pi N}(y,b)$, using the central values of our parameters for  $f_{\pi N}(y)$.}
\label{fig:bsquared}
\end{figure}

The next step is to show that the splitting functions arise  from the long-range structure of the nucleon.   Fig.~\ref{fig:brho} displays  the transverse probability distribution $2 \pi b \rho_{\pi N}(y,b)$ for several values of $y$, the momentum fraction carried by the pion.  Central values of the parameters are used.  Examination shows that the distribution is greatest for $y \approx 0.2$ and all distributions peak at a transverse distance $b \approx 0.6$ fm. This is a larger value than the 0.5 fm shown in Fig.~3 of Ref.~\cite{Strikman:2009bd}. This greater distance is caused by the use of our soft form-factors. Furthermore, Fig.~\ref{fig:bsquared}  displays the mean-square value of $b$. This distribution peaks at small $y$,  but there are long-ranged  contributions to the $\pi N$ transverse probability for all values of $y$. Strikman and Weiss \cite{Strikman:2009bd} note that for small values of $b$ the transverse distribution of pions in the nucleon is strongly dependent on form factors and cutoffs, and so argue that the pion cloud contribution can only be safely determined for large $b \ge 0.5$ fm. Our calculation  using  the experimentally-constrained  soft vertex functions shows that the pionic effects are of long-range (on the order of 1 fm) for all values of $y$.  This means that  it is not necessary to  eliminate all contributions  below 0.5 fm.  This unnecessary constraint used  in  Ref.~\cite{Strikman:2009bd} led to their Fig.~7, showing that the pion-cloud effect accounts for only for about half of the observed asymmetry.   Furthermore the long-ranged nature of the contributions verify that large values of $t$ are not important in these calculations. 

Finally,  it has been known for  a long time that the use of soft form factors (similar in range to those of the present study) leads to a convergent perturbation series~\cite{Theberge:1980ye,Thomas:1981vc,Theberge:1981mq,AlvarezEstrada:1982bx,Crawford:1983ab}. Thus the present perturbative procedure is justified.   

Having justified the model, let's turn to the observations.
The integrated asymmetry  $\bar{D}-\bar{U}$ is the difference in number of $\bar d$ and $\bar u$ quarks in the proton sea. With $\bar{D} = \int_0^1 \bar{d}(x)dx, \,\bar{U} = \int_0^1 \bar{u}(x)dx$, the asymmetry is determined from \eq{bard} and \eq{baru} as 
\bea
\bar{D}- \bar{U} ={2\over 3}\int_0^1dy f_{\pi N}(y)-{1\over 3}\int_0^1dy f_{\pi \D}(y).
\label{intasym}
\eea
The experiment E866 \cite{Towell:2001nh} measured 
 $\bar{D}-\bar{U}=0.118 \pm 0.012$.  Our splitting functions predict 
 $0.98 \leq \bar{D}-\bar{U} \leq 0.131$, in excellent agreement with the experimental result.\\
\begin{figure}
\includegraphics[width=8.0cm]{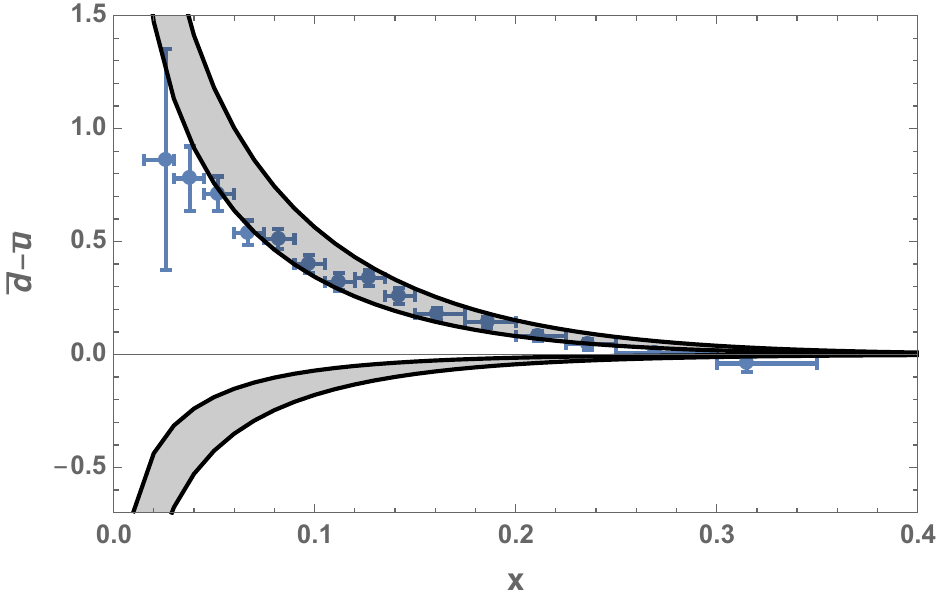}
\includegraphics[width=8.0cm]{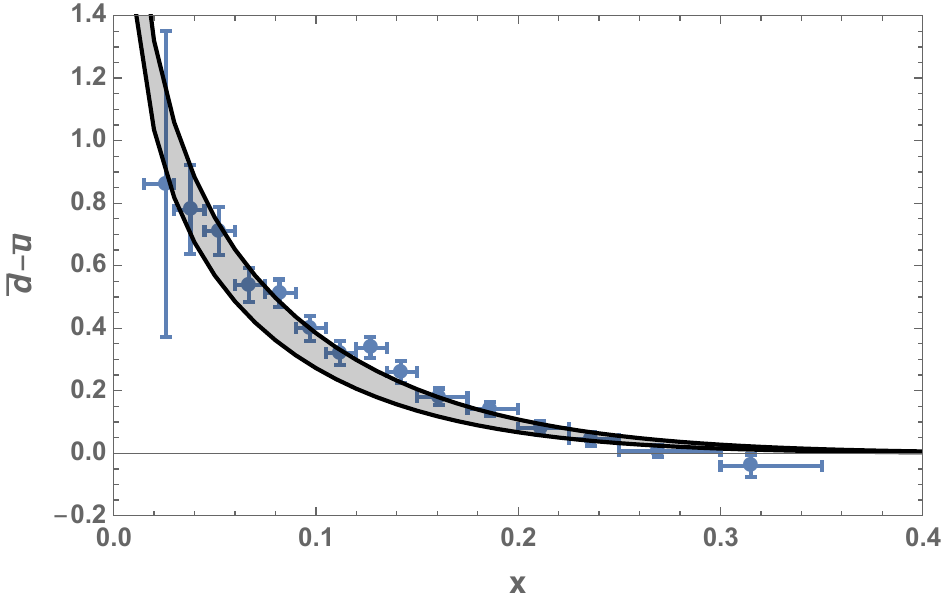}
\caption{$\bar{d}(x)- \bar{u}(x)$. Blue symbols E866 data \cite{Towell:2001nh}. The bands are computed using minimum and maximum values of the splitting functions shown in Fig.~\ref{fig:splitfcns2}. In the top panel, the upper band is for  $p \rightarrow \pi N$, the lower is for $p \rightarrow \pi \Delta$. In the lower panel, the band represents  the sum of the two contributions.}\label{fig:diff}\end{figure}
{ The computed values of  $\bar{d}(x)- \bar{u}(x)$ are compared to measurements 
 in  Fig.~\ref{fig:diff}, with bands obtained }using minimum and maximum values of the splitting functions shown in Fig.~\ref{fig:splitfcns2}. The upper band is  for $p \rightarrow \pi N$, the lower is for $p \rightarrow \pi \Delta$. The central band shows the sum of these two contributions.  Its width is narrow because of the correlation between the coupling constants: $g_{\pi \D} = r g_{\pi N}$, with $1.5 \leq r \leq 1.7$,  and the use of the same cutoff $\Lambda$ for both terms. The                               central band is a definitive prediction of the present model.  We stress that in {\it any} model, $\bar u$ and $\bar d$ are correlated so that errors in each are   partially cancelled in the ratio. 
  We find that a 15\% uncertainty  in $\bar{ d},\bar {u}$  at $x=0.3$ translates to 7\% in the ratio. 

\begin{figure}
\includegraphics[width=8.0cm]{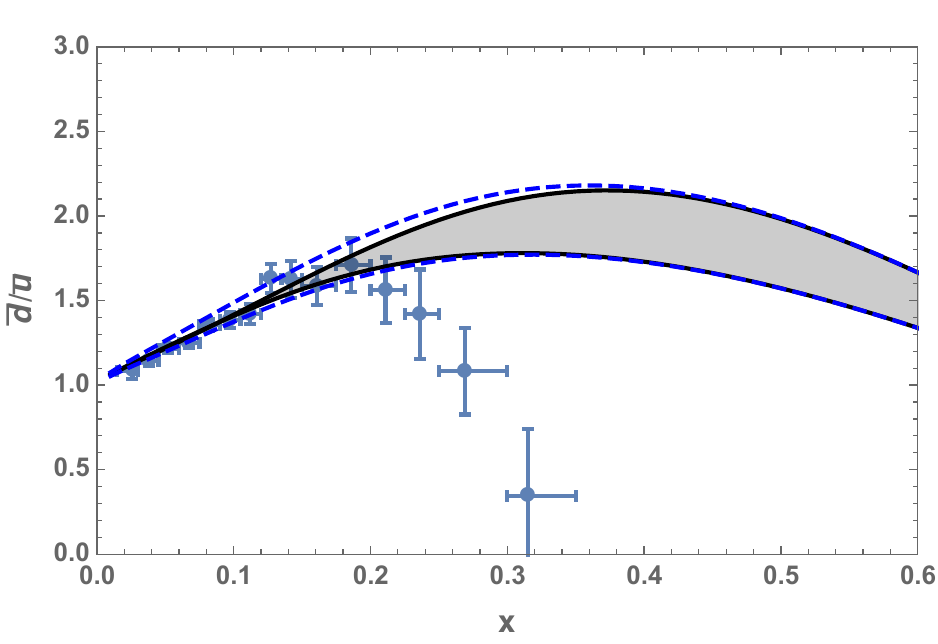}
\includegraphics[width=8.0cm]{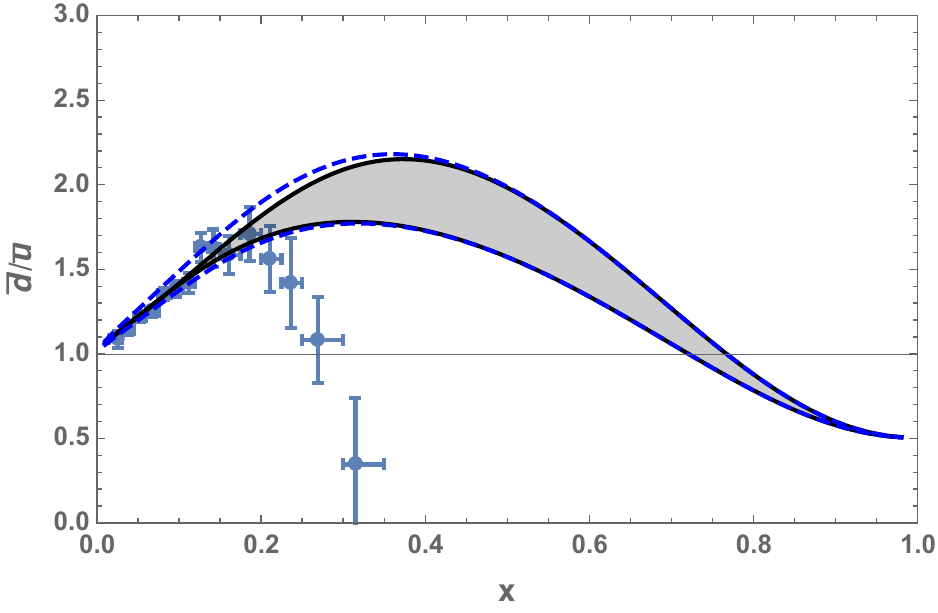}
\caption{$\bar{d}(x)/ \bar{u}(x)$ Blue symbols E866 data \cite{Towell:2001nh}. In the upper panel, the domain of the plot includes the range of $x$ covered in the SeaQuest experiment \cite{Aidala:2017ofy}.  The solid band is computed using minimum and maximum values of the splitting functions shown in Fig.~\ref{fig:splitfcns2}, using the bare sea of \cite{Szczurek:1997fr}. The dashed band includes the effects of varying the bare sea by a factor of 0.75 or 1.25. The dashed band represents our prediction for the results of  the SeaQuest experiment.In the lower panel, the domain of the the plot is extended to $x$=1.}\label{fig:ratio}

\end{figure}

The calculations of the   ratio $\bar{d}(x)/\bar{u}(x)$ are compared with   experimental data in  Fig.~\ref{fig:ratio}.  
 The results for values of $x$ less than about 0.15 arise from a combination of  pion cloud effects and the symmetric sea of the bare nucleon. 
  For larger values of $x$, terms of Fig.~2b dominate, with  the $\pi N$ contribution rising   with increasing $x$  until  $x \approx 0.4$. The ratio then drops  because of the enhancement of $\bar{u}$ (\eq{baru}) provided by   the   $\pi \Delta$ contribution, which becomes  relatively more  important as  $x$ increases.  
  
Excellent agreement with experimental data is obtained  for $x < 0.2$, but the  decrease in the ratio $\bar{d}(x)/\bar{u}(x)$ for higher values of $x$ is not reproduced. This  disagreement might  seem to rule out this model
calculation. However,  E866 is the only data set that  impacts
this quantity, and it is therefore  important to determine if this behavior is
correct. 
The displayed band predicts the results of the SeaQuest experiment, which will cover the range $0.1 \leq x \leq 0.6$ and this should definitively resolve these questions.

 A point of interest in previous literature, due to the dramatic turn-down of the data (Fig.~5), is the  limit
as $x \rightarrow 1$.  
For large  values  of $x$ the pion valence quark distributions dominate, although both $\bar{u}(x)$ and $\bar{d}(x)$ become vanishingly small. The ratio
 \begin{equation}
{\bar{d}(x)\over \bar{u}(x)}\approx {({5\over 6}f_{\pi N}+{1\over 3}f_{\pi \Delta})\otimes q^v_\pi \over ({1\over 6}f_{\pi N}+{2\over 3}f_{\pi \Delta})\otimes q^v_\pi }
\end{equation}
does not vanish,  and
  ${\bar{d}(x){}/ \bar{u}(x)}$   approaches 1/2, because of the explained greater importance of the ${\pi \Delta}$ term for  $x \rightarrow 1$.
{ This shows one mechanism that allows $\bar d/\bar u $ to drop below unity, but} it is not likely that experiments will ever reach such values of $x$. 

Some readers may be concerned that this model's  form factors produce a flavor-singlet sea $x(\bar u+\bar d)$ in excess of what is allowed by empirical parton densities that account for QCD evolution.
That this is not the case is shown in Fig.~8 that compares $x(\bar u+\bar d)$ for the present model to the NLO CT14 calculation. The resulting distributions are seen to be  well below what is allowed. The contribution of the bare sea (included in the plot of the total flavor-singlet sea), determined at $Q^2 = 4$ GeV$^2$, is also much smaller than the CT14 distribution. The effects of QCD evolution of the bare sea from $Q^2 = 4$ GeV$^2$  to $Q^2 = 54$ GeV$^2$  are to decrease the bare sea for $x > 0.14$ and increase it for $x < 0.14$ \cite{Tanabashi:2018oca}. These changes are smaller than the uncertainty bands that we used for the bare sea, and much less than the difference between CT14 (dashed line) and our calculation (solid line). Therefore our total sea does not exceed the bounds set by CT14.
\begin{figure}
\includegraphics[width=8.0cm]{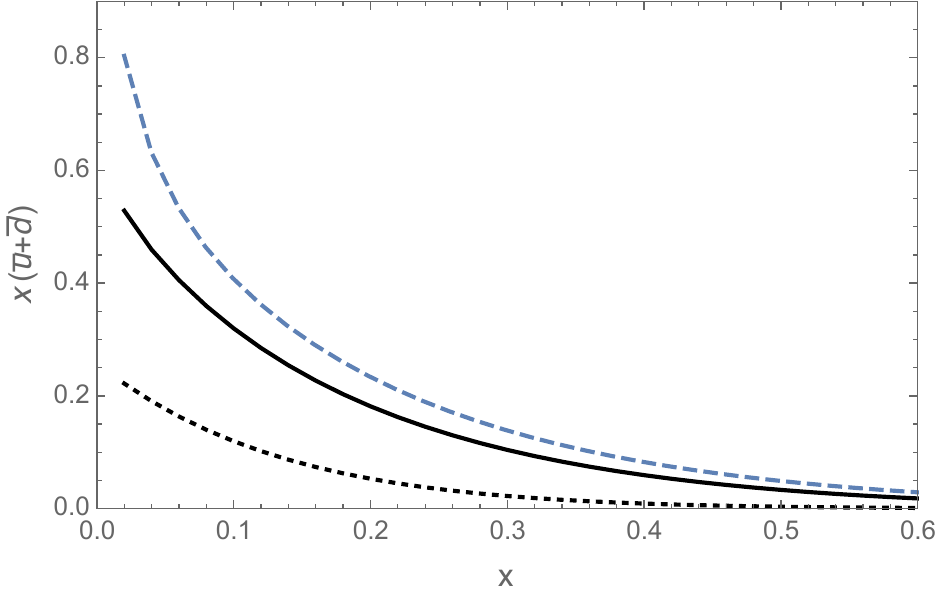}
\caption{The flavor-singlet sea ${x(\bar u}+{\bar d})$ of this model (solid line), compared to the next-to-leading order CT14 calculation (dashed line) for 
$Q^2 = 54$ GeV$^2$ . The dotted line  is the contribution of the bare sea to our model calculation. Central values of the distributions are used for all curves. } 
\end{figure}

\section{V. Summary}

 The pion-baryon form factors  of our model are essentially model-independent, and the coupling constants are reasonably well-determined. For values of $x$ greater than about 0.15, the pion cloud effects dominate. The rise and then fall of the ratio $\bar{d}/\bar{u}$ are unalterable consequences of our approach. Significantly changing any of the input parameters would cause severe disagreements with other areas of nuclear physics, and would be tantamount to changing the model. If the high$-x$  E866 results  were to be confirmed by the SeaQuest experiment, the model would be ruled out.   

 In summary, this work  presents  a chiral light-front perturbation theory calculation of the  wave function that describes  the flavor content of the nucleonic  light-quark sea. The formalism shows how to properly  obtain vertex functions 
 in a four-dimensional treatment. This allows us to  obtain results which include the effects of the  uncertainties in the input parameters,  small  enough to provide a definitive test of the pion cloud's role in the nucleon sea.  The pion cloud influence in the nucleon sea  will be ruled out if our results  were to disagree  with the eventual results of the SeaQuest experiment. 

 \section*{Acknowledgements}
The work of M.A. was supported by the Research in
Undergraduate Institutions Program of the US National Science
Foundation under Grant No. 1516105.
The work of G.A.M. was supported by the
{ USDOE}  Office of Science, Office of Nuclear
Physics under Grant No. DE-FG02-97ER41014. Support by MIT's Laboratory for Nuclear Science, USDOE Office of Science, Office of Nuclear Physics under Grant No. DE-FG02-94ER40818 is also gratefully acknowledged by G.A.M.


\begin{thebibliography}{99}
  \bibitem{Amaudruz:1991at} 
  P.~Amaudruz {\it et al.} [New Muon Collaboration],
  ``The Gottfried sum from the ratio F2(n) / F2(p),''
  Phys.\ Rev.\ Lett.\  {\bf 66}, 2712 (1991).
  
  \bibitem{Thomas:1983fh}
  A.~W.~Thomas,
  ``A Limit on the Pionic Component of the Nucleon Through SU(3) Flavor Breaking in the Sea,''
  Phys.\ Lett.\  {\bf B126}, 97 (1983).
  
   \bibitem{Henley:1990kw}
  E.~M.~Henley, G.~A.~Miller,
  ``Excess of anti-D over anti-U in the proton sea quark distribution,''
  Phys.\ Lett.\  {\bf B251}, 453-454 (1990).

 
\bibitem{Hawker:1998ty} 
  E.~A.~Hawker {\it et al.} [NuSea Collaboration],
  ``Measurement of the light anti-quark flavor asymmetry in the nucleon sea,''
  Phys.\ Rev.\ Lett.\  {\bf 80}, 3715 (1998).
  
\bibitem{Towell:2001nh} 
  R.~S.~Towell {\it et al.} [NuSea Collaboration],
  ``Improved measurement of the anti-d / anti-u asymmetry in the nucleon sea,''
  Phys.\ Rev.\ D {\bf 64}, 052002 (2001).
 
 \bibitem{Aidala:2017ofy} 
  C.~A.~Aidala {\it et al.} [SeaQuest Collaboration],
  ``The SeaQuest Spectrometer at Fermilab,''
  arXiv:1706.09990v2 [physics.ins-det].
 
\bibitem{Speth:1996pz} 
  J.~Speth and A.~W.~Thomas,
  ``Mesonic contributions to the spin and flavor structure of the nucleon,''
  Adv.\ Nucl.\ Phys.\  {\bf 24}, 83 (1997).
  
\bibitem{Chang:2014jba} 
  W.~C.~Chang and J.~C.~Peng,
  ``Flavor Structure of the Nucleon Sea,''
  Prog.\ Part.\ Nucl.\ Phys.\  {\bf 79}, 95 (2014).

\bibitem{Theberge:1980ye} 
  S.~Theberge, A.~W.~Thomas and G.~A.~Miller,
  ``The Cloudy Bag Model. 1. The (3,3) Resonance,''
  Phys.\ Rev.\ D {\bf 22}, 2838 (1980)
  Erratum: [Phys.\ Rev.\ D {\bf 23}, 2106 (1981)].
 
\bibitem{Thomas:1981vc} 
  A.~W.~Thomas, S.~Theberge and G.~A.~Miller,
  ``The Cloudy Bag Model of the Nucleon,''
  Phys.\ Rev.\ D {\bf 24}, 216 (1981).

\bibitem{Theberge:1981mq} 
  S.~Theberge, G.~A.~Miller and A.~W.~Thomas,
  ``The Cloudy Bag Model. 4. Higher Order Corrections to the Nucleon Properties,''
  Can.\ J.\ Phys.\  {\bf 60}, 59 (1982).
\bibitem{Bernard:1992qa} 
  V.~Bernard, N.~Kaiser, J.~Kambor and U.~G.~Meissner,
  ``Chiral structure of the nucleon,''
  Nucl.\ Phys.\ B {\bf 388}, 315 (1992).
\bibitem{Bernard:1995dp} 
  V.~Bernard, N.~Kaiser and U.~G.~Meissner,
  ``Chiral dynamics in nucleons and nuclei,''
  Int.\ J.\ Mod.\ Phys.\ E {\bf 4}, 193 (1995).
\bibitem{Holtmann:1996be} 
  H.~Holtmann, A.~Szczurek and J.~Speth,
  ``Flavor and spin of the proton and the meson cloud,''
  Nucl.\ Phys.\ A {\bf 596}, 631 (1996).
  
    \bibitem{Koepf:1995yh}
  W.~Koepf, L.~L.~Frankfurt, M.~Strikman,
  ``The Nucleon's virtual meson cloud and deep inelastic lepton scattering,''
  Phys.\ Rev.\  {\bf D53}, 2586-2598 (1996).

\bibitem{Strikman:2009bd}
  M.~Strikman, C.~Weiss,
  ``Chiral dynamics and partonic structure at large transverse distances,''
  Phys.\ Rev.\  {\bf D80}, 114029 (2009).
  \bibitem{Strikman:2010pu} 
  M.~Strikman and C.~Weiss,
  ``Quantifying the nucleon's pion cloud with transverse charge densities,''
  Phys.\ Rev.\ C {\bf 82}, 042201 (2010).
\bibitem{Alberg:2012wr} 
  M.~Alberg and G.~A.~Miller,
  ``Taming the Pion Cloud of the Nucleon,''
  Phys.\ Rev.\ Lett.\  {\bf 108}, 172001 (2012).
 
  \bibitem{Ji:2009jc} 
  C.~R.~Ji, W.~Melnitchouk and A.~W.~Thomas,
  ``Equivalence of pion loops in equal-time and light-front dynamics,''
  Phys.\ Rev.\ D {\bf 80}, 054018 (2009).
  
\bibitem{Burkardt:2012hk} 
  M.~Burkardt, K.~S.~Hendricks, C.~R.~Ji, W.~Melnitchouk and A.~W.~Thomas,
  ``Pion momentum distributions in the nucleon in chiral effective theory,''
  Phys.\ Rev.\ D {\bf 87}, no. 5, 056009 (2013).
  
  \bibitem{Ji:2013bca} 
  C.~R.~Ji, W.~Melnitchouk and A.~W.~Thomas,
  ``Anatomy of relativistic pion loop corrections to the electromagnetic nucleon coupling,''
  Phys.\ Rev.\ D {\bf 88}, 076005 (2013).
\bibitem{Salamu:2014pka} 
  Y.~Salamu, C.~R.~Ji, W.~Melnitchouk and P.~Wang,
  ``$\bar d - \bar u$ asymmetry in the proton in chiral effective theory,''
  Phys.\ Rev.\ Lett.\  {\bf 114}, 122001 (2015).
\bibitem{Granados:2015rra} 
  C.~Granados and C.~Weiss,
  ``Light-front representation of chiral dynamics in peripheral transverse densities,''
  JHEP {\bf 1507}, 170 (2015).
 \bibitem{Granados:2016jjl} 
  C.~Granados and C.~Weiss,
  ``Light-front representation of chiral dynamics with $\Delta$ isobar and large-N$_{c}$ relations,''
  JHEP {\bf 1606}, 075 (2016).
  
  \bibitem{Lepage:1980fj} 
  G.~P.~Lepage and S.~J.~Brodsky,
  ``Exclusive Processes in Perturbative Quantum Chromodynamics,''
  Phys.\ Rev.\ D {\bf 22}, 2157 (1980).
  
  \bibitem{Brodsky:1997de} 
  S.~J.~Brodsky, H.~C.~Pauli and S.~S.~Pinsky,
  ``Quantum chromodynamics and other field theories on the light cone,''
  Phys.\ Rept.\  {\bf 301}, 299 (1998).
   
  
\bibitem{Brodsky:2000ii} 
  S.~J.~Brodsky, D.~S.~Hwang, B.~Q.~Ma and I.~Schmidt,
  ``Light cone representation of the spin and orbital angular momentum of relativistic composite systems,''
  Nucl.\ Phys.\ B {\bf 593}, 311 (2001).
   
 \bibitem{Kovchegov:2012mbw} 
  Y.~V.~Kovchegov and E.~Levin,
 ``Quantum chromodynamics at high energy,''
  Cambridge Monogr.\ Math.\ Phys.\  {\bf 33}, 7 (2012).
  
  \bibitem{Becher:1999he} 
  T.~Becher and H.~Leutwyler,
  ``Baryon chiral perturbation theory in manifestly Lorentz invariant form,''
  Eur.\ Phys.\ J.\ C {\bf 9}, 643 (1999).

\bibitem{odgt}
  T.~R.~Hemmert, B.~R.~Holstein and N.~C.~Mukhopadhyay,
  ``N N, N Delta couplings and the quark model,''
  Phys.\ Rev.\ D\ {\bf 51}, 158  (1995);
   S.~-L.~Zhu and M.~J.~Ramsey-Musolf,
  ``The Off diagonal Goldberger-Treiman relation and its discrepancy,''
  Phys.\ Rev.\ D\ {\bf 66}, 076008  (2002).

\bibitem{Pascalutsa:1998pw} 
  V.~Pascalutsa,
  ``Quantization of an interacting spin - 3 / 2 field and the Delta isobar,''
  Phys.\ Rev.\ D {\bf 58}, 096002 (1998).
  
   \bibitem{Alberg:1999bc} 
  M.~A.~Alberg, E.~M.~Henley and G.~A.~Miller,
  ``Omega meson cloud and the proton's light anti-quark distribution,''
  Phys.\ Lett.\ B {\bf 471}, 396 (2000).
  


\bibitem{Bedaque:2002mn} 
  P.~F.~Bedaque and U.~van Kolck,
  ``Effective field theory for few nucleon systems,''
  Ann.\ Rev.\ Nucl.\ Part.\ Sci.\  {\bf 52}, 339 (2002).
 
\bibitem{Machleidt:2011zz} 
  R.~Machleidt and D.~R.~Entem,
  ``Chiral effective field theory and nuclear forces,''
  Phys.\ Rept.\  {\bf 503}, 1 (2011).

  \bibitem{Machleidt:2017vls} 
  R.~Machleidt,
  ``Historical perspective and future prospects for nuclear interactions,''
  Int.\ J.\ Mod.\ Phys.\ E {\bf 26}, no. 11, 1730005 (2017).
  
\bibitem{Yan:1973qf} 
  T.~M.~Yan,
  ``Quantum field theories in the infinite momentum frame 3. Quantization of coupled spin one fields,''
  Phys.\ Rev.\ D {\bf 7}, 1760 (1973).

\bibitem{Miller:1997cr} 
  G.~A.~Miller,
  ``Light front treatment of nuclei: Formalism and simple applications,''
  Phys.\ Rev.\ C {\bf 56}, 2789 (1997).

\bibitem{Thomas:2001kw}
  A.~W.~Thomas, W.~Weise,
  ``The Structure of the Nucleon,''
  Berlin, Germany: Wiley-VCH (2001).
  
\bibitem{Guichon:1982zk} 
  P.~A.~M.~Guichon, G.~A.~Miller and A.~W.~Thomas,
  ``The Axial Form-factor of the Nucleon and the Pion - Nucleon Vertex Function,''
  Phys.\ Lett.\  {\bf 124B}, 109 (1983).

  \bibitem{Carbonell:1998rj} 
  J.~Carbonell, B.~Desplanques, V.~A.~Karmanov and J.~F.~Mathiot,
  ``Explicitly covariant light front dynamics and relativistic few body systems,''
  Phys.\ Rept.\  {\bf 300}, 215 (1998).
  
    \bibitem{Miller:2009fc} 
  G.~A.~Miller and B.~C.~Tiburzi,
  ``The Relation Between Equal-Time and Light-Front Wave Functions,''
  Phys.\ Rev.\ C {\bf 81}, 035201 (2010).
  
\bibitem{Burkardt:2002hr} 
  M.~Burkardt,
  ``Impact parameter space interpretation for generalized parton distributions,''
  Int.\ J.\ Mod.\ Phys.\ A {\bf 18}, 173 (2003).
  
  \bibitem{arXiv:0706.3011} 
  C.~Alexandrou, G.~Koutsou, T.~.Leontiou, J.~W.~Negele and A.~Tsapalis,
  ``Axial Nucleon and Nucleon to Delta form factors and the Goldberger-Treiman Relations from Lattice QCD,''
  Phys.\ Rev.\ D\ {\bf 76}, 094511  (2007)
  [Erratum-ibid.\ D\ {\bf 80}, 099901  (2009)].
  
  \bibitem{Sullivan:1971kd}
  J.~D.~Sullivan,
  ``One pion exchange and deep inelastic electron - nucleon scattering,''
  Phys.\ Rev.\  {\bf D5}, 1732 (1972).
 
\bibitem{Aicher:2010cb} 
  M.~Aicher, A.~Schafer and W.~Vogelsang,
  ``Soft-gluon resummation and the valence parton distribution function of the pion,''
  Phys.\ Rev.\ Lett.\  {\bf 105}, 252003 (2010);
   M.~Aicher, private communication.

\bibitem{Gluck:1999xe} 
  M.~Gluck, E.~Reya and I.~Schienbein,
  ``Pionic parton distributions revisited,''
  Eur.\ Phys.\ J.\ C {\bf 10}, 313 (1999).

\bibitem{Szczurek:1997fr} 
  A.~Szczurek, V.~Uleshchenko, H.~Holtmann and J.~Speth,
  ``Production of the $W$ bosons and $Z$ bosons in the nucleon - anti-nucleon collisions and the meson cloud in the nucleon,''
  Nucl.\ Phys.\ A {\bf 624}, 495 (1997).

\bibitem{Perez:2016aol} 
  R.~N.~Perez, J.~E.~Amaro and E.~Ruiz Arriola,
  ``Precise Determination of Charge Dependent Pion-Nucleon-Nucleon Coupling Constants,''
  Phys.\ Rev.\ C {\bf 95}, 064001 (2017).
 
 \bibitem{Hill:2017wgb} 
  R.~J.~Hill, P.~Kammel, W.~J.~Marciano and A.~Sirlin,
  ``Nucleon Axial Radius and Muonic Hydrogen,''
  Rept.\ Prog.\ Phys.\  {\bf 81},  096301 (2018).
  

  \bibitem{Bernard:2001rs} 
  V.~Bernard, L.~Elouadrhiri and U.~G.~Meissner,
  ``Axial structure of the nucleon: Topical Review,''
  J.\ Phys.\ G {\bf 28}, R1 (2002).


\bibitem{Juszczak:2009qa} 
  C.~Juszczak,
  ``Running NuWro,''
  Acta Phys.\ Polon.\ B {\bf 40}, 2507 (2009)
  
\bibitem{Katori:2016yel} 
  T.~Katori and M.~Martini,
  ``Neutrino-nucleus cross sections for oscillation experiments,''
  J.\ Phys.\ G {\bf 45}, no. 1, 013001 (2018).
  
\bibitem{Nakamura:2016cnn} 
  S.~X.~Nakamura {\it et al.},
  ``Towards a Unified Model of Neutrino-Nucleus Reactions for Neutrino Oscillation Experiments,''
  Rept.\ Prog.\ Phys.\  {\bf 80}, no. 5, 056301 (2017).
\bibitem{Meyer:2016oeg} 
  A.~S.~Meyer, M.~Betancourt, R.~Gran and R.~J.~Hill,
  ``Deuterium target data for precision neutrino-nucleus cross sections,''
  Phys.\ Rev.\ D {\bf 93}, no. 11, 113015 (2016).
  
\bibitem{Bockmann:1999nu} 
  R.~Bockmann, C.~Hanhart, O.~Krehl, S.~Krewald and J.~Speth,
  ``The pi N N vertex function in a meson theoretical model,''
  Phys.\ Rev.\ C {\bf 60}, 055212 (1999).
  
\bibitem{Ericson:2000md} 
  T.~E.~O.~Ericson, B.~Loiseau and A.~W.~Thomas,
  ``Determination of the pion nucleon coupling constant and scattering lengths,''
  Phys.\ Rev.\ C {\bf 66}, 014005 (2002).


  \bibitem{AlvarezEstrada:1982bx} 
  R.~F.~Alvarez-Estrada and A.~W.~Thomas,
  ``Further Studies of Convergence in the Cloudy Bag Model,''
  J.\ Phys.\ G {\bf 9}, 161 (1983).
   \bibitem{Crawford:1983ab}
   G.~A.~Crawford and G.~A.~Miller,
   ``Convergent self-energy in the Cloudy Bag Model", Phys. \ Lett. \ {\bf  B132}, 173 (1983).

\bibitem{Tanabashi:2018oca} 
  M.~Tanabashi {\it et al.} [Particle Data Group],
  ``Review of Particle Physics,''
  Phys.\ Rev.\ D {\bf 98}, 030001 (2018).
 
   \end{thebibliography}
 \end{document}